\documentclass[amsmath,amssymb]{amsart}
\usepackage{graphicx}
\usepackage{eucal}

\pdfoutput=1 

\begin{document}

\title[Spatial evolution of tumors with successive driver mutations]{Spatial evolution of tumors\\ with successive driver mutations}

\author[Antal]{Tibor Antal$^1$}
\address{$^1$School of Mathematics, Edinburgh University, Edinburgh, United Kingdom}

\author[Krapivsky]{P. L. Krapivsky$^2$}
\address{$^2$Department of Physics, Boston University, Boston, Massachusetts, USA}

\author[Nowak]{M. A. Nowak$^3$}
\address{$^3$Program for Evolutionary Dynamics, Department of Mathematics, and Department of Organismic and Evolutionary Biology, Harvard University, Cambridge, Massachusetts, USA}

\begin{abstract}
We study the spatial evolutionary dynamics of solid tumors as they obtain additional driver mutations. We start with a cancer clone that expands uniformly in three dimensions giving rise to a spherical shape. We assume that cell division occurs on the surface of the growing tumor. Each cell division has a chance to give rise to a mutation that activates an additional driver gene. The resulting clone has an enhanced growth rate, which generates a local ensemble of faster growing cells, thereby distorting the spherical shape of the tumor. We derive analytic formulas for the geometric boundary that separates the original cancer clone from the new mutant as well as the expanding frontier of the new mutant. The total number of original  cancer cells converges to a constant as time goes to infinity, because this clone becomes enveloped by mutants. We derive formulas for the abundance and diversity of additional driver mutations as function of time. Our model is semi-deterministic: the spatial growth of the various cancer clones follows deterministic equations, but the arrival of a new mutant is a stochastic event. 
\end{abstract}

\maketitle

\section{Introduction}

Cancer arises when somatic cells receive multiple mutations that enhance their net reproductive rate \cite{vogelstein98}. Tumors contain 35 to 70 genetic alterations that change protein sequences \cite{vogelstein13}. The vast majority of those mutations are passengers that do not confer a selective growth advantage. A small subset, however, are driver mutations that promote tumorigenesis. In the human genome about 135 genes are known that can function as drivers when mutated (either by point mutation, insertion, deletion or amplification). Driver mutations affect pathways that regulate cell survival, proliferation and genome maintenance. Any one tumor contains between 2 to 8  driver mutations \cite{vogelstein13}. In this paper we study the accumulation of such drivers in a spatial model of tumor growth. 

Mathematical models of cancer evolution have studied the age incidence of cancers \cite{moolgavkar81}, the effect of tissue geometry and chromosomal instability \cite{nowak02,komarova03} on cancer initiation; the inactivation of tumor suppressor genes \cite{nowak04}, the accumulation of driver and passenger mutations in expanding tumors \cite{beerenwinkel07,bozic10}; the molecular clock of cancer \cite{yachida10} and the emergence of resistance to cancer therapy \cite{komarova05,michor05,iwasa06,komarova06,bozic13}.  

Modeling the genetic evolution of cancer has been predominantly performed in the homogeneous setting. This is an obvious idealization, especially for solid tumors, but it greatly simplifies the mathematical analysis.
The homogeneous setting allows researches to focus on the temporal dynamics. It provides a useful theoretical laboratory to probe the efficacy of drug combinations. A more faithful spatial modeling is necessary \cite{bryne06,roose07} for understanding tumor invasion and metastasis \cite{baraldi13}, and efforts in this direction are growing. Previous spatial models mainly focus on the evolution of already existing types of cells in space. Most models are either continuum mathematical models consisting of partial differential equations  \cite{roose07,sherratt92} or discrete cell population models using cellular automata-type computer simulations  \cite{Torq11}. Simulations are often performed at cell levels, incorporating cell movement and different cell types, and are either lattice based or off-lattice \cite{roose07,Torq11,matteis13}. When using partial differential equations to describe the density of different cell types in space, the boundary of the tumor is also evolving (free boundary problem) \cite{friedman06,bryne97,cui01}.

Here we break new ground by developing a geometric approach for the accumulation of driver mutations in spatially expanding tumors. The spatial inhomogeneity of tumors and the spatial distribution of genetic mutations has been studied in recent experimental and theoretical works \cite{yachida10,sottoriva12}.
Since different mutations are present in different spatial regions of the tumor, spatial inhomogeneity is relevant for choosing the optimal targeted drug therapy for patients.  In this paper we are mainly interested in the evolving shape of the tumor and its interplay with the onset of successive driver mutations. We deliberately simplify the model as much as possible, while keeping the key features, namely the spatial growth and the competition between different mutants. The goal is to eventually apply spatial tumor modeling of the accumulation of driver mutations to problems which were recently analyzed in the idealized framework of space-less cancer, that is in well mixed population of cells \cite{beerenwinkel07,bozic10,durrett10}, as well as to other problems which can only be formulated in the spatial framework.

Our model is reminiscent of the pioneering lattice model of cancer which incorporates mutation \cite{williams72,bramson80,bramson81}. In contrast to this earlier work, we assume that mutations occur only on the surface of the growing tumor. Furthermore, we assume that the spatial expansion is deterministic. Only mutational events are stochastic. We also mention a few more recent related studies. In Ref.~\cite{martens11}, the accumulation of many successive driver mutations was studied by computer simulations on a two-dimensional lattice. It was found that space makes the arrival of new driver mutations slower than in a well mixed population. Since including both space and mutation make models quite complex, one usually resorts to simulation results and approximations. Conversely, in \cite{komarova06,komarova13,nowak03} analytic results are derived for one-dimensional tissue geometry  and in \cite{durrett12} for the accumulation of neutral mutations in any dimension.

Our model has two basic ingredients: stochastic nucleation of new mutants and deterministic growth of existing cell types. Nucleation and growth are ubiquitous natural phenomena, and our model overlaps with classical models of such processes. Perhaps the closest connection is with the polynuclear growth model of crystals (see \cite{Evans93} for a review). Similar models have been used in cosmology (see \cite{Kleban11} for a review). The contrasting features of our model is the nucleation on the surface of the growing tumor and differences in the growth rates; in other applications nucleation events usually happen in the bulk and growth rates are equal. For example, in cosmological applications \cite{Kleban11} cosmic bubbles grow at a speed of light. In our model a mutation activating a driver gene leads to enhanced growth rate leading to the distortion of the spherical shape of the original tumor. We analyze in detail the simplest case of the competition between the original cancer clone  and one mutant clone. We establish analytical formulas for the boundary separating the clones, and determine the time when the mutant clone envelopes the original cancer clone which thereby ceases to grow any further.


\section{Results}

In our model, cells only proliferate on the surface of the tumor. Inside the tumor, cells are non dividing, hence there are no evolutionary dynamics there. This assumption is plausible for early stages of tumor progression, where only tumor cells close to the surface can get enough oxygen or other nutritions to divide. A typical tumor developing in vivo has most of its cell proliferation constrained to the border \cite{bru03,drasko05}, which suggests that cell surface diffusion is the main mechanism responsible for growth in any type of tumor. At later stages of tumor progression, when angiogenesis starts to work, this assumption may no longer be valid, but there can always be interior regions with low supply of nutrients and oxygen and low activity of cell division.

The dynamics of our model is given by the growth rate on the surface of the tumor, and by the arrival rate of new driver mutations. Without mutations, the original tumor grows spherically \cite{chaplain01,drasko05,ciarletta13}. Since cell divisions only occur on the surface of the tumor, mutations can only arrive there, at a constant rate per unit surface area and unit time. We include in this mutation rate the survival probability of mutant clones. In other words, we are only tracking mutants that survive. Since we assume that these mutants have selective advantage over the original tumor, the mutant clones keep spreading. By setting the length scale and the time scale, we set the speed of the original tumor growth and the mutation rate to one. Hence without mutations, the original tumor is a ball of radius $t$ at time $t$. We are mostly interested in the three-dimensional case, but we also present a few basic results for two dimensions.

Now we have to specify the tumor growth in the presence of advantageous mutants. Each point on the surface is characterized by a growth speed, corresponding to different mutant types. The surface of the tumor then grows in the normal direction at the rate of the local growth speed. 

In the simplest case we consider two types of cells: (i) the initiating cancer cell with growth rate one; and (ii) a mutant cancer cell with growth rate $v>1$. The mutant cell arises by activation of an additional driver. The surface of the tumor either belongs to a mutant clone or the original tumor. A point at a distance $dt$ from the surface of the tumor will be occupied by a mutant clone $dt$ times later, if a mutant can reach that point earlier than a non-mutant (see a more detailed description later). Since a mutant clone grows a distance $vdt$ during this time, a surface location will be occupied by a mutant if there is a mutant clone on the surface within a distance $\beta dt$, with
\begin{equation}
 \beta = \sqrt{v^2-1}
\end{equation}
Hence the mutant area on the surface is expanding, with the boundary moving at constant speed $\beta$.

Due to the simplicity of the model, there is only a single parameter $v$ (or equivalently $\beta$). We have achieved this by rescaling the length-scale and the time scale. The dependence of the results on the detailed parameters is discussed below.


\subsection{Shape of mutant clones}

\begin{figure}[t]
\begin{center}
\includegraphics[width=0.9\textwidth]{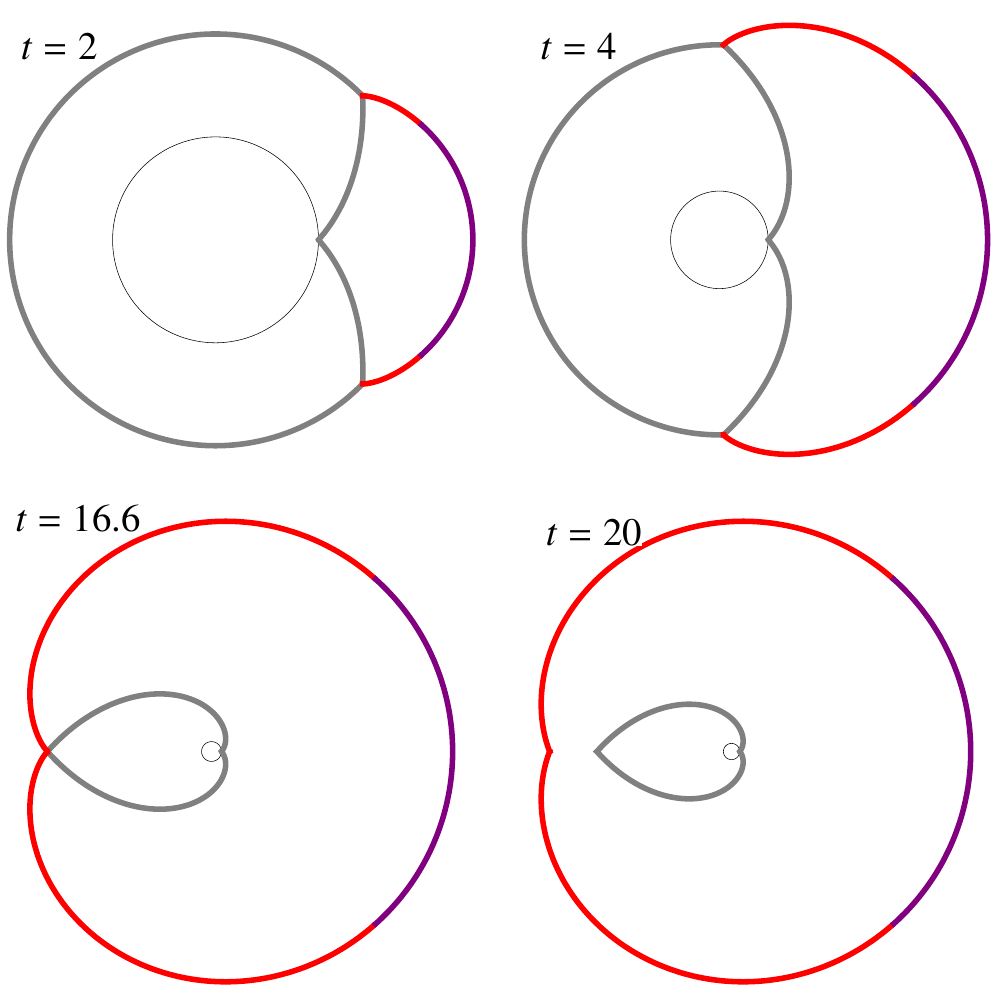}
\caption{\small Slices of a tumor with a single mutant clone at times $t=2,4,16.6,20$. The slice is along any plane which goes through the initial points of the original and the mutant clones. The original tumor is initiated at the origin, and the mutant is initiated at $t=1$ at $(1,0,0)$ and has fitness $v=1.5$. The tumor at the time of mutant initiation is drawn with thin black line at each stage to show the length scale. The boundaries of the original tumor is depicted by gray, and the outer boundary of the mutant clone is red and purple: referring to the different functional forms of the curves. On the lower left picture the original tumor is captured by the mutant clone at $t_\mathrm{c}=e^{\pi/\beta}=16.6087\dots$, and on the lower right one the mutant overgrows the enclosed original tumor.}
\label{Fig:earlier}
\end{center}
\end{figure}

Let us describe first the shape of the mutant clones. Let us focus on the shape of a single clone, as they all look identical. At time $t=0$ the original tumor starts growing spherically. Let us initiate a single mutant clone at time $t=1$ from point $(x,y,z)=(1,0,0)$ in cartesian coordinates. At this point the original tumor covers a ball of radius one around the origin. Since the tumor stays rotationally symmetric around the $x$ axes, we describe here a two dimensional cut through the $(x,y)$ plane. Since the shape of the tumor is a revolution body around the $x$ axes, we only give the boundaries for $y\ge 0$. For a mutant clone initiated at spherical coordinates $(r_0, \theta_0, \phi_0)$ the shape is the same as the one initiated at $(r_0=1, \theta_0=0, \phi_0=0)$, but with space and time stretched by $r_0$ and rotated by $\theta_0$, $\phi_0$.

\begin{figure}[t]
\begin{center}
\includegraphics[width=.8\textwidth]{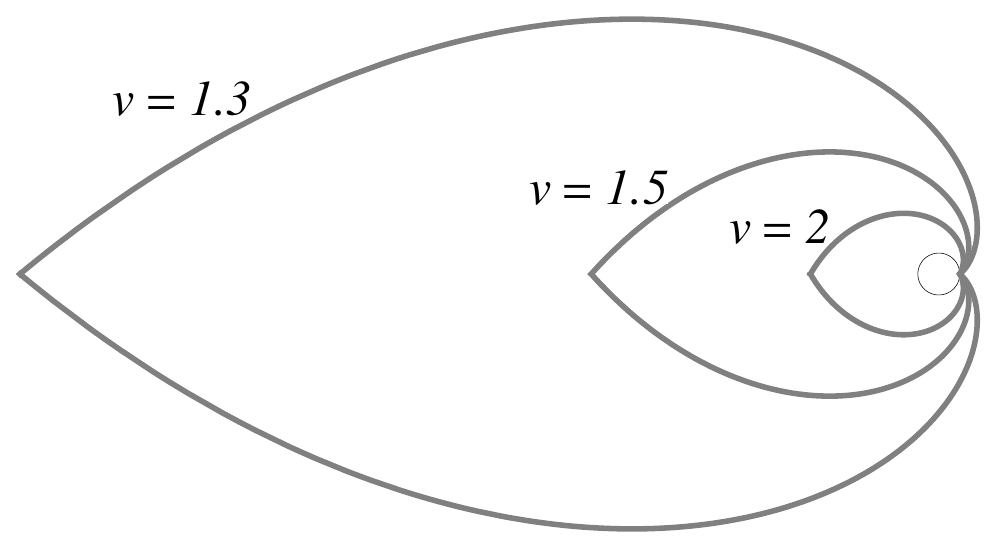}
\caption{\small The final "barnacle" shape of the original tumor after it has been captured by the mutant clone at different fitness values $v=1.3, 1.5, 2$ of the mutant. The mutant is always initiated at $t=1$, and the capture takes place at $t_\mathrm{c}=e^{\pi/\beta}=43.9052, 16.6087, 6.13371$ respectively for the different fitness values. The black circle in the middle represents the original tumor at the initiation of the mutant clone $t=1$.}
\label{Fig:barnacle}
\end{center}
\end{figure}

In the two dimensional cut through the $(x,y)$ plane, the shape of the original tumor at time $t$ has generally two parts: the boundary between the original tumor and the mutant, which in polar coordinates is $r(\theta)=e^{\theta/\beta}$, or in cartesian coordinates
\begin{equation}
\label{between}
\begin{split}
x(\theta)&=e^{\theta/\beta} \cos \theta\\ 
y(\theta)&=e^{\theta/\beta} \sin \theta
\end{split}
\end{equation}
for $0\le \theta\le \theta_0$, where 
\begin{equation}
\label{theta_0}
 \theta_0 = \beta \log t
\end{equation}
For $\theta>\theta_0$, the original tumor is a sphere of radius $t$ around the origin.

The mutant at time $t$ is separated from the original tumor by the boundary given by \eqref{between}, and the mutant's outer limits are given by two segments. The middle part is a sphere around $(1,0)$ with radius $v(t-1)$
\begin{equation}
\begin{split}
x(\theta)&= 1+v(t-1)\cos \theta\\ 
y(\theta)&= v(t-1)\sin \theta
\end{split}
\end{equation}
for $0\le \theta\le \arccos 1/v$, and the outer part next to the original tumor is given by
\begin{equation}
\label{outer}
\begin{split}
x(\theta) &= [(1-\beta)e^{\theta /\beta}+\beta t]\cos\theta -(t-e^{\theta /\beta})\sin\theta\\
y(\theta) &= [(1-\beta)e^{\theta /\beta}+\beta t]\sin\theta -(t-e^{\theta /\beta})\cos\theta
\end{split}
\end{equation}
for $0 \le \theta\le \beta \log t$. Note that $\theta$ is only a parameter here, and not a polar coordinate. 
Rotating these curves around the $x$ axes we obtain the surfaces to the tumor clones.

\begin{figure}[t]
\begin{center}
\includegraphics[width=\textwidth]{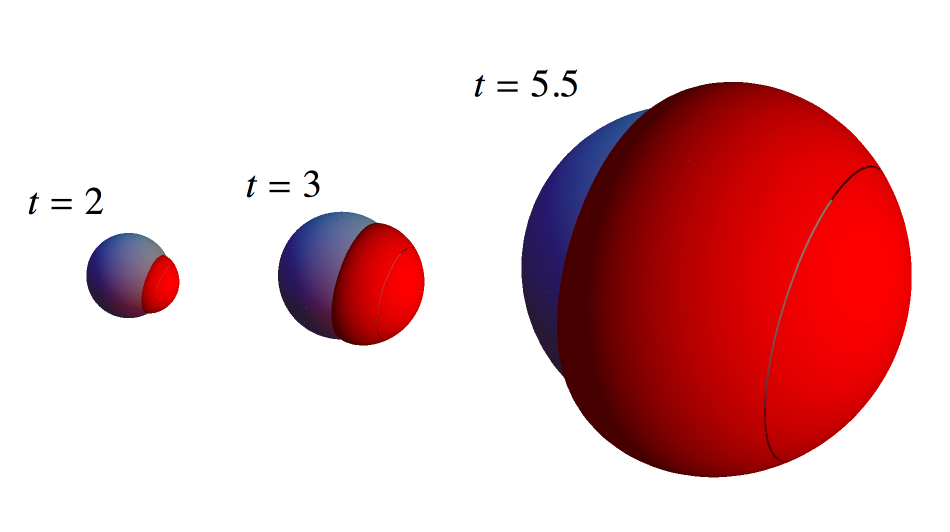}
\caption{\small Shape of the tumor with a single mutant clone at times $t=2,3$ and 5.5. The mutant is initiated at $t=1$ and has fitness $v=1.5$. The boundaries of the original tumor is depicted by gray, and that of the mutant clone by red.}
\label{Fig:3dsingle}
\end{center}
\end{figure}

\begin{figure}[t]
\begin{center}
\includegraphics[width=0.6\textwidth]{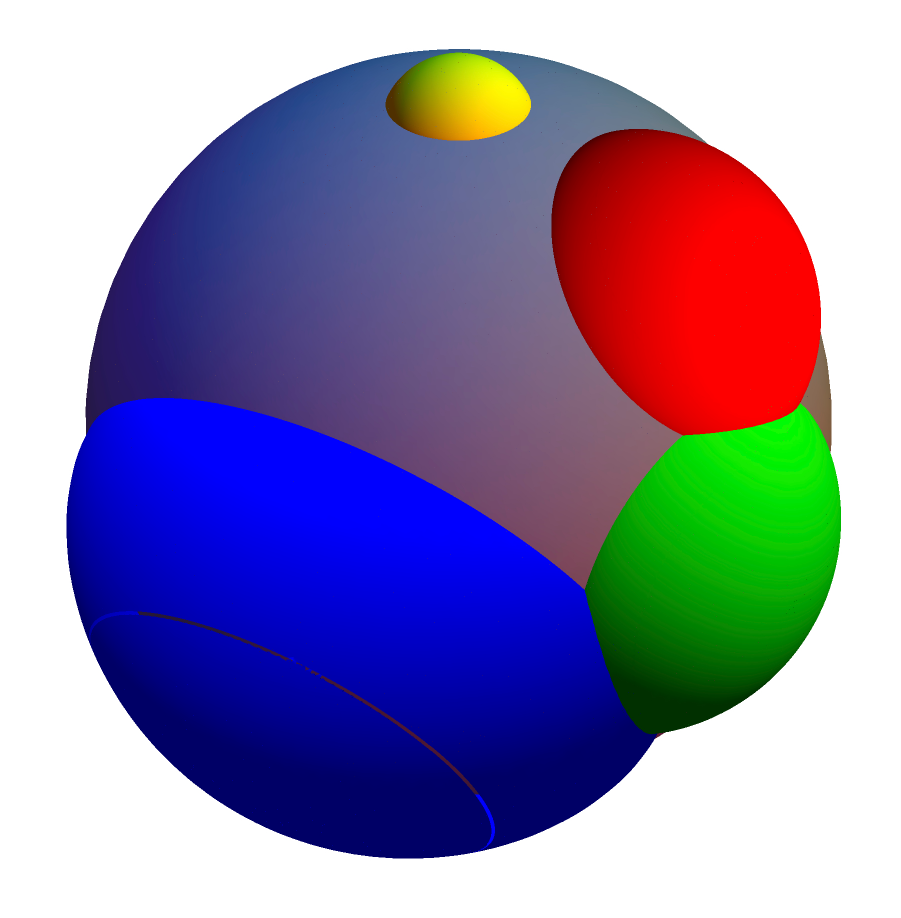}
\caption{\small Shape of the tumor with many mutant clones arrived at different times with different fitness values. Mutant clones are initiated stochastically at constant rate on the surface of the original tumor, and then they grow deterministically at constant rate. The growth rates of the mutant clones were chosen randomly between 1 and 2 for this illustration.}
\label{Fig:3dmulty}
\end{center}
\end{figure}

The boundary on the surface of the tumor between the original tumor and the mutant are at an angle $\theta_0=\beta \log t$ with the $x$ axes from the origin. When this angle becomes $\pi$, the original tumor is completely covered by the mutant, which happens at time
\begin{equation}
 t_\mathrm{c} = e^{\pi/\beta}
\end{equation}
After this time the original tumor ceases to grow any further, and its final volume is
\begin{equation}
\label{Vfinal}
V_\mathrm{c}(\beta) = \frac{2\pi}{3}\, \frac{\beta^2}{\beta^2+9} \left( 1+e^{3\pi/\beta} \right)
\end{equation}
A two dimensional cut of a single mutant clone is depicted on Fig~\ref{Fig:earlier} for several time points, and on Fig~\ref{Fig:barnacle} the final shape of the original tumor is shown after its capture by a single mutant. On Fig~\ref{Fig:3dsingle} the tumor with a single mutant clone is depicted in three dimension, and on Fig.~\ref{Fig:3dmulty} a tumor with multiple mutant clones is drawn for illustration.

\subsection{Total volume of original tumor clone}

Now we allow several mutations to arrive at the tumor, and we are interested in the total volume of the mutant clones and original tumor clone. Since outside of the ball of radius $t$ all tumor cells are mutants, the question is the ratio of mutant clones inside the ball of radius $t$. That is also the probability that a random point inside this ball is a mutant. There can be many mutant clones and they can touch each other too. We assume, however, that no successive mutations arrive inside mutant clones, or at least that those ``second order" clones stay confined inside their originator mutant clone.

A time $t$, a random point at distance $r$ from the origin (that is on the sphere of radius $r$), with $r\le t$, is non-mutant with probability
\begin{equation}
 W_r = e^{-b(\beta)r^3}
\end{equation}
with
$$
b(\beta) = \frac{2\pi}{3}\, \frac{\beta^2}{\beta^2+9} \left( 1+e^{-3\pi/\beta} \right)
$$
A random point in the tumor within a ball of radius $r$, with $r\le t$, is non-mutant with probability
\begin{equation}
 W_{\le r} 
 = \frac{1-e^{-b(\beta)r^3}}{b(\beta)r^3}
\end{equation}
This is the fraction of non-mutant volume in the tumor within radius $r$.
The total non-mutated tumor volume tends to a constant for large times
\begin{equation}
 \lim_{t\to\infty} \frac{4\pi}{3} t^3 W_{\le t} = \frac{4\pi}{3b(\beta)}
\end{equation}

\subsection{Time till first mutant clone}

Let's denote the number of mutant clones by $N_t$ at time $t$.
We can give an exact result for the probability of no mutant clones at time $t$, which is also the probability that the arrival time $T$ of the first mutant clone is greater than $t$.
Since the total rate of arrival of mutants till time $t$ is just the volume of the sphere
\begin{equation}
\label{lambda}
 \Lambda_t = \frac{4\pi t^3}{3}
\end{equation}
hence
$$
 P(N_t=0) = P(T>t)= e^{-\Lambda_t} = e^{-4\pi t^3/3}
$$
that is the first mutant arrives according to the density function
$$
 f_T(t) = 4 \pi  t^2 e^{-4 \pi  t^3/3}
$$
Consequently, the first mutant arrives after a mean time with variance 
$$
ET= \frac{\Gamma(1/3)}{6^{2/3} \pi^{1/3}} \approx 0.55396, \quad 
\mathrm{Var} T = \frac{6 \Gamma(2/3)-\Gamma(1/3)^2}{6^{4/3} \pi^{2/3}} \approx 0.0405358
$$
Since the original tumor grows at rate one, the first mutant clone appears also at distance $T$ from the origin. That is it appears on average at distance $ET\approx 0.55396$.

If we wait long enough, a mutant will appear with probability one. The probability that there are no further mutations from the original tumor, so the original tumor has a final barnacle shape, is
$$
\frac{9+\beta^2}{\beta^2}   \frac{2}{1+ e^{3\pi/\beta}}
$$
This probability is quite small for realistic relative speeds; it is around $0.36\%$ for $v=1.5$, and around $3.5\%$ for $v=2$, although it approaches one as $v\to\infty$.

\subsection{Number of different clones}

If we allow subsequent mutations within mutant clones, and assume that all mutation rates are one, what is the total number of clones $N_t$ at time $t$? 
Since with mutations the shape of the tumor becomes very irregular, it is hard to give an exact expression for larger values of $N_t$. But let us approximate the tumor as a ball of radius $t$ at time $t$, which is a not too bad approximation if all the fitnesses are sufficiently similar.
the total number of mutants in this approximation is $N_t\sim \mathrm{Poisson}(\Lambda_t)$, that is
$$
 P(N_t=n) \approx \frac{\Lambda_t^n}{n!} e^{-\Lambda_t}
$$
The mean number of clones and its variance is
$$
 EN_t \approx \mathrm{Var} N_t \approx \Lambda_t
$$

\subsection{Reduction of parameters}
\label{parameters}

Our basic model has only a single parameter, $v$, denoting the relative growth rate of mutant clones.
But this is the consequence of a reduction of parameters, which we discuss now. 
Let us measure time in days, and distance in cm. In general we have the following parameters describing the system.
The surface of the original tumor grows in the normal direction at rate $\mathcal{V}_0$, and mutations arrive at the surface of the tumor at rate $\mathcal{U}$ per unit time and unit surface area. The mutant clone growth at rate $\mathcal{V}_1$.
Let us define the new unit length and time as
$$
 L_0 = \left( \frac{\mathcal{V}_0}{\mathcal{U}}\right)^{1/3} \quad   
 T_0 = (\mathcal{UV}_0^2)^{-1/3} = \frac{L_0}{\mathcal{V}_0}
$$
Measuring length and time in these new units, the original clone grows at rate one, and mutations arrive at rate one. 
The speed of the fronts and mutation rates per surface area might be directly accessible experimentally. 
Having obtained the unit length and time $L_0, T_0$ for the tumor, all results of the paper could be used when replacing time with $t\to t/T_0$ and all lengths with $l\to l/L_0$.
The scaled speed of the mutant clone is 
$$
v=\mathcal{V}_1/\mathcal{V}_0
$$ 
which is the only parameter of the scaled model.

We can obtain some estimates for the values of the above parameters as follows.
In our model the original tumor grows only on the surface as a sphere. Starting from a single cell it reaches volume $V_T$ in time $T$. In the scaled coordinates the tumor is just a ball of radius scaled time, but here we include explicitly the scaling for the space and time units as given above to obtain
$$
 \frac{V_T}{L_0^3} = \frac{4\pi}{3} \left( \frac{T}{T_0} \right)^{1/3}
$$
Equivalently, we can rewrite this expression as
$$
 V_T = \frac{4\pi}{3} (\mathcal{V}_0 T)^3
$$
This gives an estimate for the growth rate
$$
 \mathcal{V}_0 = \frac{1}{T} \left( \frac{3 V_T}{4\pi} \right)^{1/3}
$$ 

We estimate the surface mutation rate from the number of driver clones found in a tumor. Our expression for the mean number of clones given in Eq.~\eqref{lambda} is
$$
 \Lambda_T = \frac{4\pi}{3} \left( \frac{T}{T_0} \right)^{1/3} = \frac{V_T}{L_0^3} 
 = V_T \frac{\mathcal{U}}{\mathcal{V}_0}
$$
which then leads to the estimate
$$
 \mathcal{U} = \frac{\mathcal{V}_0 \Lambda_T}{V_T}
$$

From the above formulas we can obtain an order estimate for our parameters.
We expect a tumor of $V_T \approx 1-10 ~\mathrm{cm}^3$ after 5 to 10 years of growth (so $T\approx 5-10\times 365$ day), and we expect of the order of $\Lambda_T\approx 1-10$ driver clones \cite{vogelstein13,bozic10}. Note that we expect more clones in larger tumors, so roughly 
$$
  \frac{\mathcal{U}}{\mathcal{V}_0} = \frac{\Lambda_T}{V_T}\approx \frac{1}{\mathrm{cm}^3}.
$$ 
This leads to the estimates
$$
 \mathcal{V}_0 \approx 10^{-3} - 10^{-4} \frac{\mathrm{cm}}{\mathrm{day}}
 \quad \quad
 \mathcal{U} \approx 10^{-3} - 10^{-4} \frac{1}{\mathrm{cm}^2 \mathrm{day}}
$$

Finally, let us estimate the relative speed of the mutant clone $v=\mathcal{V}_1/\mathcal{V}_0$. In \cite{bozic10} it was estimated that the birth rate of cells with $k$ driver mutations is larger by $sk$ than their death rate (that is their fitness is $sk$), with $s$ being $0.005$. If the original clone has $k$ driver mutations, the mutant clone is expected to have $k+1$ mutations. We assume that the speed of a clone is proportional to its fitness advantage, and since everything else is assumed to be the same in the clones, the relative speed of the mutant clone becomes
$$
 v = \frac{k+1}{k}
$$
Since $k$ is typically an integer between 1 and 8 \cite{vogelstein13,bozic10}, the speed is $1< v\le 2$.
This is the only parameter of the scaled model.


\section{Derivations}

The tumor occupies a subset of the $d$-dimensional space $T\subset \mathbf{R}^d$, and each point has a fitness $f:T\to\mathbf{R}^+$. The tumor can only grow at the surface in the normal direction each point at rate $f(\cdot)$ (wherever the surface is differentiable). Hence to obtain the shape of the tumor an infinitesimally small time $dt$ later, draw a ball of radius $f(\cdot) dt$ around each point on the surface of the tumor, and the outer envelope of the union of these balls becomes the new surface. 

If a mutant of fitness $v>1$ is initiated at a point on a locally flat surface of the original tumor of fitness one, then $dt$ times later the mutant occupies a sector of radius $v dt$ and half angle $\phi=\arccos(1/v)$, while the original tumor progressed a distance $dt$ and it is around the mutant sector. The angle between the surface of the original and the mutant is $\pi-\phi$, and it stays constant during the evolution. After the initiation the boundary of the mutant clone keeps moving at speed $\beta=\sqrt{v^2-1}$ on the surface of the original tumor.  

\subsection{Shape of clones}

The original tumor is initiated at the origin at time $t=0$. Let us focus on the shape of a mutant clone initiated at $t=r_0$ at the cartesian point $(r_0,0)$ in $d=2$ or $(r_0,0,0)$ in $d=3$. Since the tangential speed of the boundary of the mutant clone on the surface (that is at distance $t$) is a constant $\beta$, the shape of the clone and the original tumor stay rotationally symmetric around the $x$ axes. Hence it is sufficient to describe the shape of the tumor in two dimensions, and for $y\ge 0$.

\begin{figure}
\begin{center}
\includegraphics[width=0.3\textwidth]{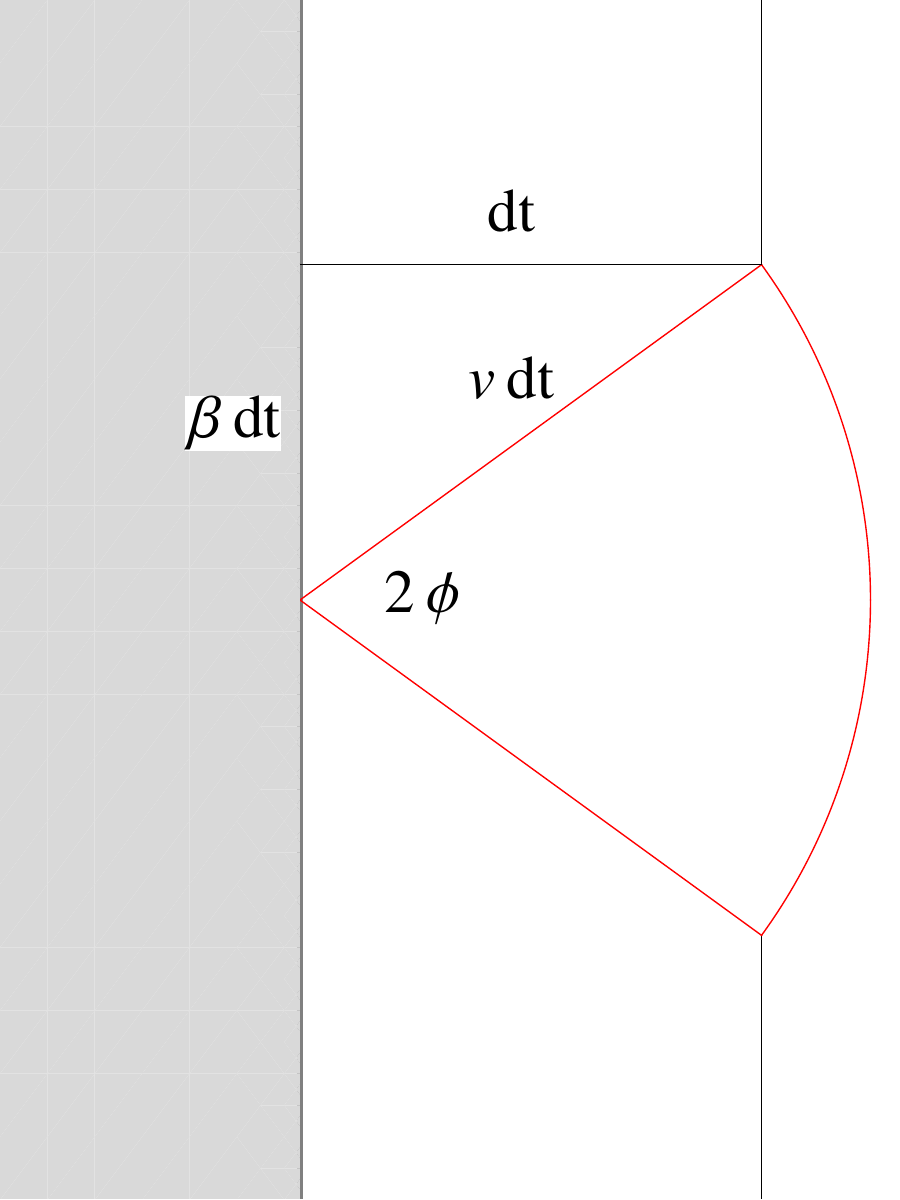}
\hspace{1cm}
\includegraphics[width=0.3\textwidth]{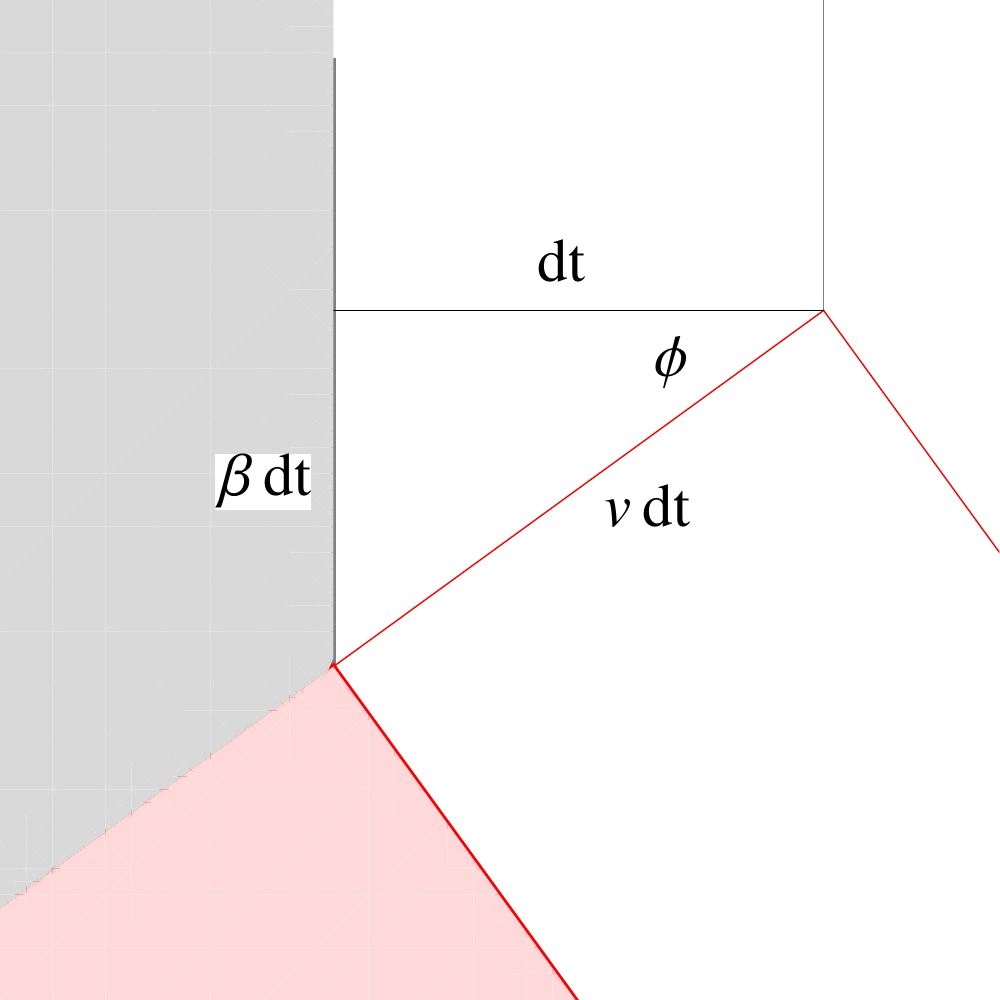}
\caption{\small Illustration for the spreading of the mutant clone. Only a two dimensional cut is shown. On the left panel the initiation of a mutant clone is captured. A tiny segment of the surface of the original clone is almost flat and depicted by a shaded grey region. For a small time interval $dt$ later the surface is composed by a circular arc and straight segments. On the right panel the evolution of the surface of the tumor is shown for later times. Note that the angle $\phi$ stays constant during the process.}
\label{Fig:growth}
\end{center}
\end{figure}

\begin{figure}
\begin{center}
\includegraphics[width=0.7\textwidth]{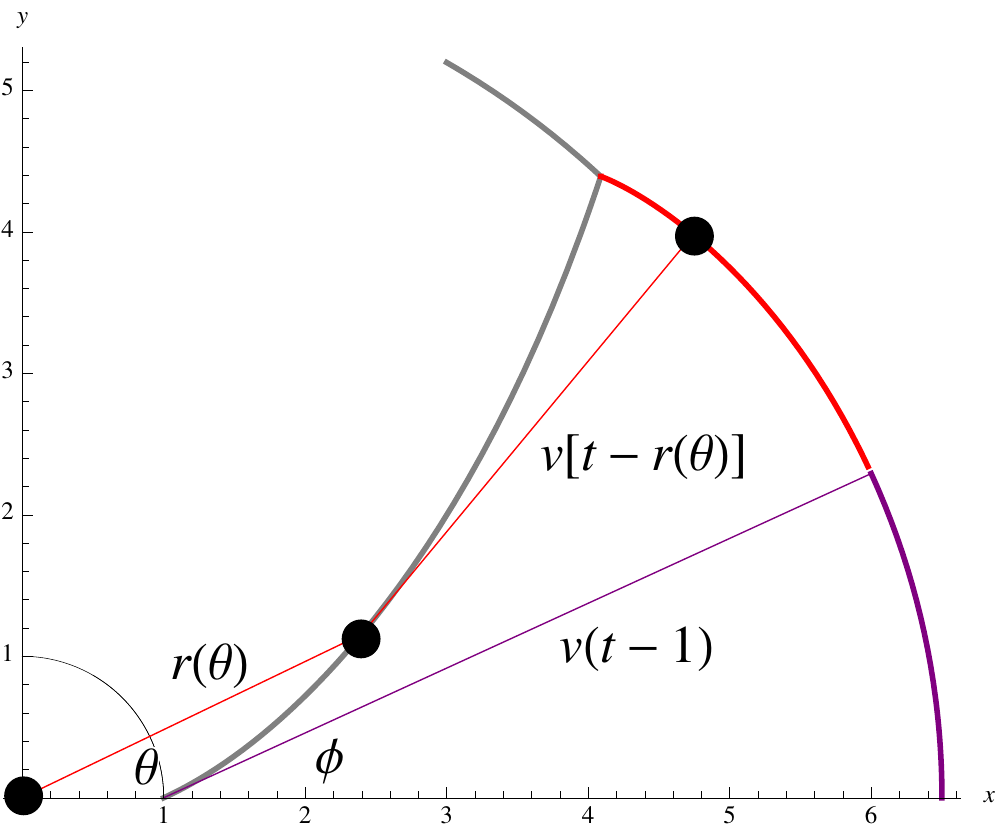}
\caption{\small Illustration for calculating the shape of the mutant clone. Only a two dimensional cut is shown. The black dot at the origin is the center of the original tumor. The mutant clone is initiated at $(1,0)$. The purple segment can be reached directly from the initial mutant position, and hence its outer boundary is a circle of radius $v(t-1)$. The thick grey line indicates the boundary of the original tumor, and the curve between the original tumor and the mutant clone is given by the parametric curve $r(\theta)$. The top black dot represents a general point on the red curve (outer boundary of themutant clone). It is reached by the mutant clone originating from the middle black dot in time $t-r(\theta)$ at speed $v$.}
\label{Fig:illust1}
\end{center}
\end{figure}

Let us use polar coordinates $(r,\theta)$ for now. The growth mechanism of the clone is explained on Fig.~\ref{Fig:growth}. A small time $dt$ after initiation the mutant clone occupies a circular segment of radius $vdt$, and angle $2\phi$ with $\phi=\arccos(1/v)$. This arc is at an angle $\pi-\phi$ with surface of the original clone, and this angle stays constant during the evolution. The boundary between the mutant and the original clone keeps moving at constant speed $\beta=\sqrt{v^2-1}$, as can be seen on Fig.~\ref{Fig:growth}. Hence the boundary of a mutant clone initiated at $(r_0, 0)$ at time $t=r_0$ is described by the differential equation
$$
 r \theta'(r) = \beta
$$
with solution
$$
 \theta(r) = \beta \log \frac{r}{r_0}
$$
or equivalently
\begin{equation}
\label{between2}
 r(\theta) = r_0 e^{\theta/\beta}
\end{equation}
for $0\le\theta\le\theta_0$, where $\theta_0=\beta\log t/r_0$, and we measure $\theta$ from the $x$ axes. This is equivalent to \eqref{between} for $r_0=1$. 

The boundary between the original tumor and the mutant becomes a closed curve at time $t_\mathrm{c}=r_0 e^{\pi/\beta}$, after which time the original tumor ceases to grow. Its final volume is calculated later. For earlier times, $t<t_\mathrm{c}$, the outer boundary of the original tumor is a sphere of radius $t$, for $\theta_0<\theta<\pi$. These boundaries, as well as the initial boundary of the original tumor (the unit circle $r=1$) are shown on Fig.~\ref{Fig:earlier} as red curves. 

The boundary of the mutant clone contains the boundary with the original tumor given by \eqref{between2} and two other pieces corresponding to the outer boundary of the mutant. The first one is a circle (green curve on Fig.~\ref{Fig:earlier}) centered at the seed of the mutant clone, i.e. at $(r_0,0)$, with radius $R=v[t-r_0]$. The opening half-angle $\phi$ of this part of the circle is found by computing the inclination angle between curve \eqref{between2} and the $x$ axis. One obtains $\phi=\arccos(1/v)$. 

To determine the remaining part of the boundary (blue curve on Fig.~\ref{Fig:earlier}) one draws straight lines in the tangential direction from each point of the curve \eqref{between2} as it is illustrated on Fig.~\ref{Fig:illust1}. The angle between this tangential and the $x$ axes is $\phi$ at any point of $r(\theta)$.
If we draw the tangential from the point given by polar coordinates $(r(\theta),\theta)$, the mutant clone has still time $t-r$ to grow, so the boundary will be at
\begin{equation*}
\begin{split}
x(\theta) &= r(\theta)\cos\theta + v[t-r(\theta)]\cos[\theta+\arccos(1/v)]\\
y(\theta) &= r(\theta)\sin\theta + v[t-r(\theta)]\sin[\theta+\arccos(1/v)]
\end{split}
\end{equation*}
which can be rewritten as 
\begin{equation*}
\begin{split}
x(\theta) &= [(1-\beta)r(\theta)+\beta t]\cos\theta -[t-r(\theta)]\sin\theta\\
y(\theta) &= [(1-\beta)r(\theta)+\beta t]\sin\theta -[t-r(\theta)]\cos\theta
\end{split}
\end{equation*}
If we now rescale both space and time by $r_0$, we recover the shape of a clone initiated at $(1,0,0)$, as given in \eqref{outer}.

Let us compute the area $A_\mathrm{cover}$ covered by the original tumor at the moment of capture. Using \eqref{between2} for $r_0=1$ we get
\begin{equation*}
A_\mathrm{c}=\int_0^\pi d\theta\,r^2(\theta)
=\frac{\beta}{2}\left(e^{2\pi/\beta}-1\right)
\end{equation*}
Similarly, in three dimensions, the volume of the original tumor at capture is given by
\begin{equation*}
V_\mathrm{c} = \frac{2\pi}{3}\int_0^\pi d\theta\,\sin\theta\, r^3(\theta)
= \frac{2\pi}{3}\,\frac{\beta^2}{\beta^2+9}\left(e^{3\pi/\beta}+1\right)
\end{equation*}
as announced in \eqref{Vfinal}.
For general $r_0$ these formulas are multiplied by $r_0^d$.

\subsection{Many mutations for $d=2$}

What fraction of the tumor is mutated at time $t$? 
Point $(r,0)$ is covered by a mutant initiated at $(r_0, \theta_0)$ if this initial point is on or within the boundaries
$$
 r_0 = r e^{-|\theta_0|/\beta}
$$
with $-\pi\le\theta_0\le\pi$. Mutations arrive as an inhomogeneous Poisson process, hence we need the total rate of arrival for such a mutation is this region
$$
 A = 2 \int_0^\pi \frac{r_0(\theta)^2}{2} d\theta = a(\beta) t^2, \quad a(\beta) = \frac{\beta}{2} \left( 1-e^{-2\pi/\beta} \right)
$$
Hence the probability of no mutant at distance $r$ in the tumor is 
$$
 W_r = e^{-a(\beta)r^2}
$$

More formally, let $w$ be a function $w:\mathbf{R^d}\to\mathbf{N}$ counting the number of subsequent mutations present at a given point in the tumor. Inside the original tumor clone $w(.)=0$, it is one at mutant clones arisen from the original tumor, and $k+1$ for mutant clones arisen from mutant clones with $w(\cdot)=k$. Let $R$ be a uniform random vector within a ball of radius $t$. Hence we just calculated the probability 
$$
 P[w(R)=0\big||R|=r] =  W_r = e^{-a(\beta)r^2}
$$

The probability that a (uniformly picked) random point in the tumor of ball $r$ is not mutated, that is the fraction of non-mutated tumor is
\begin{equation*}
\begin{split}
 W_{\le r} &= P[w(R)=0\big||R|\le r] = E[P(w(R)=0|R)\big||R|\le r]\\ 
 &= \frac{1}{\pi r^2} \int_0^r  e^{-a(\beta)r'^2} 2\pi r' dr' = \frac{1-e^{-a(\beta)r^2}}{a(\beta)r^2}\\
\end{split}
\end{equation*}
Interestingly, the non-mutated tumor mass tends to a constant for large times
$$
 \lim_{t\to\infty} \pi t^2 W_{\le t} = \frac{\pi}{a(\beta)}
$$

\subsection{Many mutations for $d=3$}

The calculation is similar in 3 dimensions. Here, the boundary of points which cover $(r,\theta=0,\phi=0)$ is given by the same expression $r_0(\theta)$ as in 2 dimensions, and the total rate of mutants arriving in this region equals to its volume, which is
$$
 V = \frac{2\pi}{3} \int_0^\pi r_0^3(\theta) \sin(\theta) d\theta = b(\beta) t^3, \quad 
 b(\beta) = \frac{2\pi}{3}\, \frac{\beta^2}{\beta^2+9} \left( 1+e^{-3\pi/\beta} \right)
$$
At time $t$, a random point at distance $r$ is non-mutant with probability
$$
 W_r=P(w(R)=0\big||R|=r) = e^{-V} = e^{-b(\beta)r^3}
$$
and a random point in a ball of radius $r$ is non-mutant with probability
\begin{equation*}
\begin{split}
 W_{\le r} &= P[w(R)=0\big||R|\le r] = E[P(w(R)=0|R)\big||R|\le r]\\ 
 &=  \frac{3}{4\pi r^3} \int_0^r  e^{-b(\beta)r'^3} 4\pi r'^2 dr' 
 = \frac{1-e^{-b(\beta)r^3}}{b(\beta)r^3}
\end{split}
\end{equation*}
As before, the non-mutated tumor mass tends to a constant for large times
$$§
 \lim_{t\to\infty} \frac{4\pi}{3} t^3  P[w(R)=0\big||R|\le t] = \frac{4\pi}{3b(\beta)}
$$

\subsection{Probability of a single mutant clone}

The probability of having no mutant clone till time $t$ goes to zero faster than exponential in time. Recall that $N_t$ is the total number of mutations raised either from the original clone, or from any mutant clones by time $t$. Conversely, let $N_{1,t}$ be the  number of clones initiated only from the original tumor by time $t$. We are interested in the eventual number of such clones $N_1=\lim_{t\to\infty} N_{1,t}$. (Note that this variable is finite with probability one, since $P(N_1=0)=0$, and the original tumor stops growing a finite time interval after the first mutant clone was initiated). 

What is the probability that there is only a single mutant clone from the original clone, that is $P(N_1=1)$? In that case we could observe the final barnacle shape of the original tumor. The first mutant clone appears at a random time $T$, and at distance $T$ from the origin. Conditioning on this time, there are no further mutations with probability
$$
 P(N_1=1 | T=t) = e^{-[V_\mathrm{c}(\beta)-4\pi/3] t^3}
$$  
since $[V_\mathrm{c}(\beta)-4\pi/3] t^3$ is the total rate of production of the second mutant.
Now taking the average over the initiation time
\begin{equation*}
\begin{split}
 P(N_1=1) &= E P(N_1=1 | T) = 4\pi \int_0^\infty t^2 e^{-V_\mathrm{c}(\beta)t^3} dt\\
 &= \frac{4\pi}{3V_\mathrm{c}} = 2 \left( 1+\frac{9}{\beta^2} \right)  \left( 1+ e^{3\pi/\beta} \right)^{-1}
\end{split}
\end{equation*}
This function monotone grows from zero to one with $0\le\beta\le\infty$.

We can also calculate the probability distribution of the time of the second mutation. Let $T_1$ be the time of the first mutation (which is finite with probability one), and let $T_2$ be the time of the second mutation with $T_2>T_1$, which is finite with probability $1-P(N_1=1)$. As before, we can write the conditional probability
$$
 P(T_2/T_1>\tau| T_1=t) = e^{-[V(\tau)-4\pi/3]t^3}
$$
where $1\le\tau\le t_\mathrm{c}=e^{\pi/\beta}$, and
$$
 V(\tau) = 2 \pi \frac{ \beta ^2+\left(\beta ^2+9\right) \tau^3
  +3 \tau^3 [\beta  \sin (\beta \log \tau)+3 \cos (\beta  \log \tau)]}{3 \left(\beta ^2+9\right)}
$$
is the volume of the original tumor. 
The simplest way to obtain this volume is from its derivative $dV/d\tau=2\pi\tau^2[1+\cos(\beta \log \tau)]$, which is the surface of a sector with half cone angle $\pi-\theta_0=\pi-\beta\log \tau$. 
Now averaging over the arrival time of the first mutant clone we obtain
$$
P(T_2/T_1>\tau) = E P(T_2/T_1>\tau| T_1) =  4\pi \int_0^\infty t^2 e^{-V(\tau)t^3} dt
= \frac{4 \pi}{3 V(\tau)} 
$$
for $1\le \tau\le e^{\pi/\beta}$. Of course, $P(T_2/T_1>1)=1$, and 
$P(T_2/T_1>\tau)=P(N_1=1) = 4\pi/(3 V_\mathrm{c})$ for $\tau\ge t_\mathrm{c}=e^{\pi/\beta}$, which corresponds to no second mutation.

\section{Discussion}

%

Many mathematical models of cancer evolution are based on the assumption of well-mixed populations. This homogeneous setting represents  a reasonable framework for the modeling of liquid tumors, but in solid tumors the effects of spatial structure can be important. Here the reliance on well-mixed cell populations is mostly caused by the better mathematical tractability of that simple framework. 

In this paper, we have developed a model of cancer which describes both spatial and temporal evolution and accounts for mutations that activate additional driver genes, leading to enhanced proliferation rates of cancer cells. Our model depends on very few parameters. In the simplest case of one mutant,  there is only a single parameter, $v$, denoting the ratio of the growth rates of the mutant clone and the initiating cancer clone. Even in this setting the emerging behavior is rich. For example, given enough time we observe the inevitable capture of the initial clone by the mutant. Hence the initial clone grows to a fixed size as time goes to infinity. The capture time, however, is much larger than the naive estimate would suggest. This finding correlates with the general conclusion emerging from other studies, see e.g. \cite{martens11}, namely that spatial structure reduces the rate of cancer progression.

Throughout this paper we assumed that if successive mutant clones are initiated inside mutant clones then they stay confined in the parental mutant clone.
But what happens if inside a clone of fitness $v_{i}$ a new mutant arrives with fitness $v_{i+1}>v_i$? The new clone's boundary has a tangential speed on the surface of the parent clone given by
$\beta_{i+1} = \sqrt{v_{i+1}^2-v_{i}^2}$. Note that it can be smaller than $\beta_i$ for a more fit mutant $v_{i+1}>v_i$.
Since the first clone eventually covers the parental clone and becomes asymptotically circular, a newly arriving clone eventually covers the previous clone.

Our model is semi-deterministic -- the spatial growth of the tumor is deterministic, while the birth of new mutants is stochastic. The former feature simplifies the analysis. The rules of the dynamics are isotropic: in isolation a mutant clone exhibits a spherical growth. Yet the stochasticity of the arrival of mutant clones and the strong interaction between the initial clone and mutant clones, and also between different types of mutant clones, results in highly anisotropic shapes. 

One of the main virtues of the model is its simplicity; we can derive exact results describing the basic behavior of the model. This simplicity is encouraging to pursue further extension of the model. It would be interesting to study the effect of random growth rates for each mutant clone, the dynamics of new mutant clones arising within mutant clones, and the time it takes to accumulate  several additional driver mutations \cite{beerenwinkel07,jones08,bozic10,durrett10,yachida10,martens11} in a spatial setting.

\section*{Acknowledgments}

Support from the John Templeton Foundation and the program for 
Foundational Questions in Evolutionary Biology is  gratefully acknowledged (T.A. and M.A.N.).

\end{document}